\documentclass[lettersize,journal]{IEEEtran}
\usepackage[utf8]{inputenc}
\usepackage{amsmath,amsfonts}
\usepackage{algorithmic}
\usepackage{array}
\usepackage[caption=false,font=normalsize,labelfont=sf,textfont=sf]{subfig}
\usepackage{textcomp}
\usepackage{stfloats}
\usepackage{url}
\usepackage{verbatim}
\usepackage{graphicx}
\usepackage{color}
\usepackage{bm}
\usepackage[numbers,sort&compress]{natbib}

\makeatletter

\newcommand{\Rmnum}[1]{\expandafter\@slowromancap\romannumeral #1@}
\makeatother
\def\BibTeX{{\rm B\kern-.05em{\sc i\kern-.025em b}\kern-.08em
    T\kern-.1667em\lower.7ex\hbox{E}\kern-.125emX}}
\usepackage{balance}
\begin{document}
\title{mmAlert: mmWave Link Blockage Prediction via Passive Sensing}
\author{Chao Yu, Yifei Sun, Yan Luo and Rui Wang}

\maketitle

\begin{abstract}
In this letter, the mmAlert system, predicting millimeter wave (mmWave) link blockage during data communication, is elaborated and demonstrated. The passive sensing method is adopted for mobile blocker detection, where two receive beams with separated radio frequency (RF) chains are equipped at the data communication receiver. One receive beam is aligned to the direction of line-of-sight (LoS) path, and the other one periodically sweeps the region close to the LoS path. By comparing the signals received by the above two beams, the Doppler frequencies of the signal scattered from the mobile blocker can be detected. Furthermore, by tracking the Doppler frequencies and the angle-of-arrivals (AoAs) of the scattered signals, the trajectory of the mobile blocker can be estimated, such that the potential link blockage can be predicted by assuming consistent mobile velocity. It is demonstrated via experiments that the mmAlert system can always detect the motions of the walking person close to the LoS path, and predict 90\% of the LoS blockage with sensing time of 1.4 seconds.
\end{abstract}


\section{Introduction}
\IEEEPARstart{D}{ue} to millimeter-level wavelength, millimeter wave (mmWave) communications are highly sensitive to the blockage of line-of-sight (LoS) path, which would lead to the link disconnection. Therefore, a robust mmWave communication system should be able to predict the LoS blockage, and prepare communication link along non-line-of-sight (NLOS) path in advance.

There have been a number of works demonstrating the blockage prediction of mmWave links via out-of-band sensors \cite{Vision, Radar, Lidar}. For example, a vision-assisted active blockage prediction scheme was proposed in \cite{Vision}, where a machine learning algorithm was introduced to forecast blockage according to the pictures taken by camera. In \cite{Radar} and \cite{Lidar}, mmWave radar and light detection and ranging (LiDAR) devices were used to monitor the surrounding mobile blockers, respectively.
On the other hand, there are also a number of works exploiting the in-band channel information in blockage prediction\cite{sub-6g1,sub-6G2,Diffraction}. For example, it was shown in \cite{sub-6g1,sub-6G2} that the channel state information (CSI) of the sub-6G band could be utilized to forecast the blockage of the mmWave links in a dual-band communication system. It was demonstrated in \cite{Diffraction} that due to the signal diffraction, a sudden increment of received signal strength indicator (RSSI) could be observed when a  mmWave link was about to be blocked. However, the abovementioned CSI-based methods \cite{sub-6g1,sub-6G2} cannot directly provide the motion information of mobile blockers and thus lead to large false alarm probability. Note that a mobile blocker around the signal path may not always lead to blockage. Moreover, the methods based on signal diffraction may not be able to provide sufficient response time for backup link preparation.

In fact, passive sensing is a promising approach to facilitate blocker detection during data communication with half-duplexing transceivers. In \cite{Car}, by exploiting the long-term evolution (LTE) downlink signal, the trajectory of a target car could be estimated by passive sensing with at most 2 m estimation error. In \cite{human-tracking}, the WiFi signal was utilized in passive sensing. It was demonstrated that human targets could be detected and tracked by jointly estimating the range, Doppler frequencies and angle of arrivals (AoAs) of the scattered signal. Moreover, it was shown in \cite{human-sensing} that the WiFi-based passive sensing was able to detect human motions, such as walking, waving hand and even breathing, behind the wall. Recently, a mmWave-based passive sensing system was developed in \cite{passive-sensing}, where different hand gestures could be distinguished with high accuracy.

Exploiting the passive sensing techniques, a mmWave blockage prediction system, namely mmAlert, is proposed in this paper. Specifically, a mmWave communication system capable of mobile blocker tracking is developed, where the receiver with two radio frequency (RF) chains could perform data receiving and blocker sensing simultaneously. By comparing the signals received via the two receive beams, the Doppler effect raised by the potential mobile blocker can be detected. Note that not all the blocker motions close to the LoS path could lead to link blockage, an estimation method is also proposed to track the trajectory of mobile blockers, such that harmful blocker motion can be detected and the false alarm probability can be supposed.

The remainder of this paper is organized as follows. In Section $\mathrm{\uppercase\expandafter{\romannumeral2}}$, an overview of the mmAlert system is provided. In Section $\mathrm{\uppercase\expandafter{\romannumeral3}}$, the signal processing for passive sensing is introduced. In Section $\mathrm{\uppercase\expandafter{\romannumeral4}}$, the algorithm for blockage prediction is proposed. The experiment results and discussion are provided in Section $\mathrm{\uppercase\expandafter{\romannumeral5}}$.  Finally, the conclusion is drawn in Section $\mathrm{\uppercase\expandafter{\romannumeral6}}$.

\section{Overview of mmAlert System}
\begin{figure}[htbp]
	\centering
	\includegraphics[width=1\columnwidth]{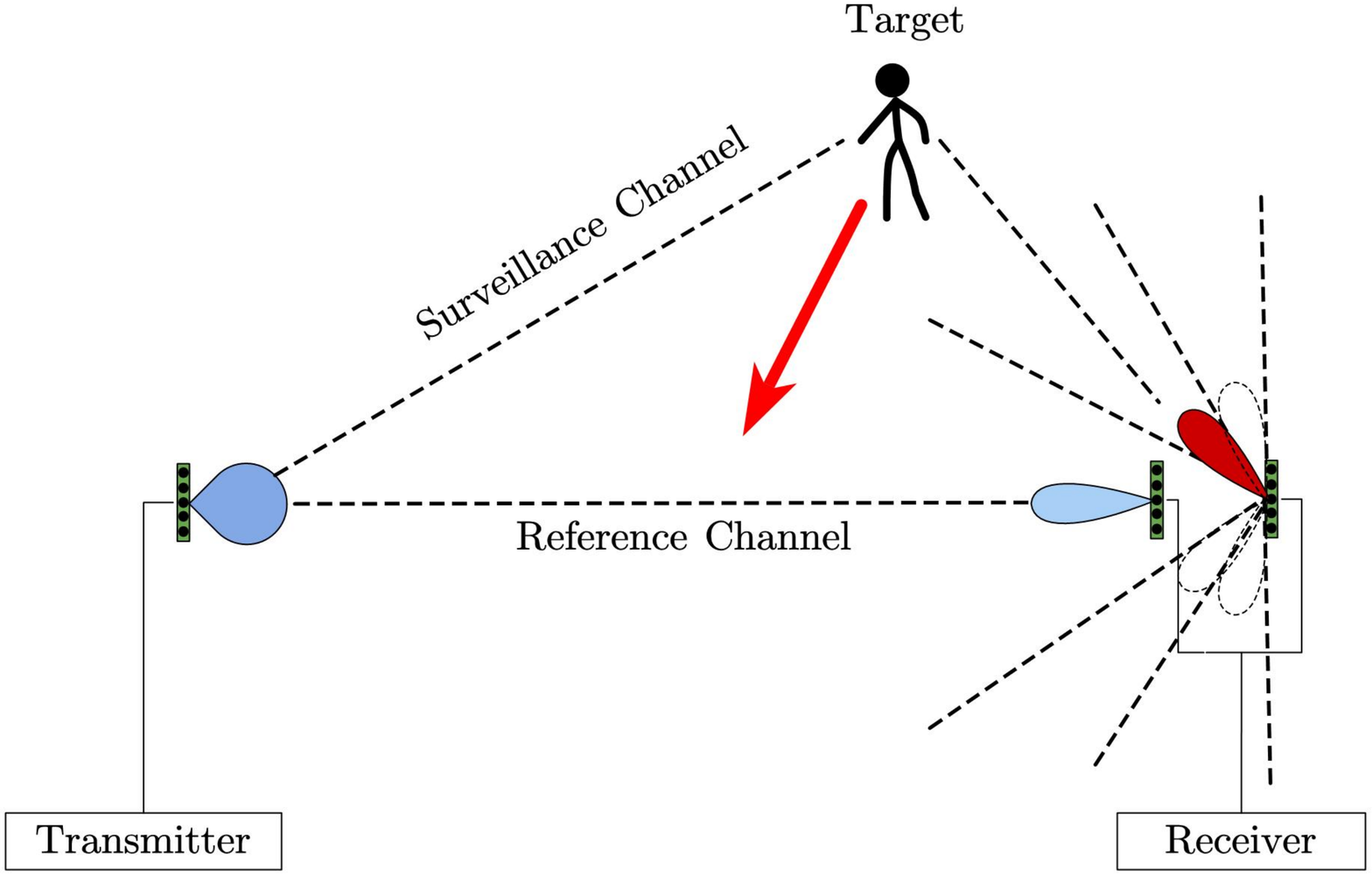}
	\caption{The overview of mmAlert system.}
	\label{fig:Passive Sensing}
\end{figure}

The mmAltert is deployed with a mmWave communication system to predict the potential LoS link blockage due to mobile blockers, e.g., walking people. 
This system is illustrated in Fig. \ref{fig:Passive Sensing}. At least two RF chains are equipped at the mmWave receiver, such that the communication and passive sensing of potential blockage can be performed simultaneously with half-duplexing transceivers. 
Specifically, the mmWave transmitter delivers an information-bearing signal via a transmission beam, while two receive beams, namely the reference beam and surveillance beam, are adopted at the mmWave receiver. 
Both the transmission and the reference beam are aligned with the LoS path, such that a high signal-to-noise ratio (SNR) can be achieved. 
The surveillance beam sweeps the region close to the LoS path periodically to detect the potential mobile blocker. 
Signals of both receive RF chains can be used jointly to retrieve the transmission message. 
Moreover, by comparing the signals of both RF chains, the Doppler effect due to mobile blockers can be detected. 
Note that not all the motions of the mobile blocker close to the LoS path would block the link, the successive detection is applied to estiamte its trajectory, such that the probability of false alarm can be significantly suppressed.

\section{Signal Processing for Passive Sensing}
\label{section:signal processing}
\subsection{Signal Model}
To simultaneously communicate with the receiver and illuminate potential blockers around the LoS, the transmitter array steers a wide beam along the LoS path. At the receiver, the narrow reference beam is steered toward the LoS direction, while the surveillance beam is periodically switched among $\mathrm{M}$ directions, denoted as $\Phi = \{\varphi_{1},\varphi_{2},\ldots,\varphi_{M}\}$. Let $\mathrm{T}_\mathrm{b}$ be the sensing duration of the surveillance beam in one direction and the time duration of one sweeping period is $\mathrm{T}_\mathrm{d} = \mathrm{M} \times \mathrm{T}_\mathrm{b}$.

Without loss of generality, we consider the signal processing when the surveillance beam is at the direction $\varphi_{m}$ of one sweeping period, say the i-th period. Denote the transmitted signal as $s_\mathrm{i,m}(t)$, then the signal received by the reference beam can be written as
\begin{align}
	y_{\mathrm{r}}^{\mathrm{i,m}}(t)
	=
	\alpha_{\mathrm{r}}s_{\mathrm{i,m}}(t-\tau_{\mathrm{r}})
	+
	n_{\mathrm{r}}^{\mathrm{i,m}}(t),
\end{align}
where $\alpha_{\mathrm{r}}$ and $\tau_{\mathrm{r}}$ denote the complex gain and delay of the LoS path, respectively, and $n_{\mathrm{r}}^{\mathrm{i,m}}(t)$ denotes the noise.

The received signal of the surveillance beam, denoted as $y_{\mathrm{s}}^{\mathrm{i},\mathrm{m}}(t)$, consists of the echo signals scattered from the mobile blocker and surrounding static scattering clusters. Thus,
\begin{equation}
\begin{aligned}
	y_{\mathrm{s}}^{\mathrm{i,m}}(t)
	=
	&\alpha_{\mathrm{i,m}}^{\mathrm{tar}}s_\mathrm{i,m}(t-\tau_{\mathrm{s,i}}^{\mathrm{tar}})e^{-j2\pi f_{\mathrm{d,i}}^{\mathrm{tar}}t}\\
	&+
	\sum_{\ell=1}^{L}\alpha_{\mathrm{i,m}}^{\ell}s_\mathrm{i,m}(t-\tau_{\mathrm{s,i}}^{\ell})
	+
	n_{\mathrm{s}}^{\mathrm{i,m}}(t),
\end{aligned}
\end{equation}
where $L$ is the number of static paths in the surveillance channel, and $n_{\mathrm{s}}^{\mathrm{i,m}}(t)$ is the noise in the surveillance channel. $\alpha_{\mathrm{i,m}}^{\mathrm{tar}}$, $\tau_{\mathrm{s,i}}^{\mathrm{tar}}$ and $f_{\mathrm{d,i}}^{\mathrm{tar}}$ denote the complex gain, delay and Doppler frequency of the scattered path from the blocker, respectively.  Similarly, $\alpha_{\mathrm{i,m}}^{\ell}$ and $\tau_{\mathrm{s}}^{\ell}$ denote the complex gain and delay of the $\ell$-th static path, respectively. 

The received signals from surveillance and reference beams are sampled at the baseband with a period $T_{s}$,  which can be expressed by
\begin{align}
	y_\mathrm{s}^{\mathrm{i},\mathrm{m}}[n]=y_\mathrm{s}^{\mathrm{i},\mathrm{m}}(nT_s),
\end{align}
and 
\begin{align}
	y_{r}^{\mathrm{i},\mathrm{m}}[n]=y_{r}^{\mathrm{i},\mathrm{m}}(nT_s),
\end{align}
respectively.
As a remark notice that there is usually strong signal leakage from the LoS path in the received signal $y_\mathrm{s}^{\mathrm{i,m}}$, which can be canceled via the least-square-based (LS-based) clutter cancellation as elaborated in \cite{tan2005passive}. 

\subsection{Doppler and AoA Estimation}

The cross-ambiguity function (CAF) for the estimation of Doppler frequency is given by
\begin{align}
	R^{\mathrm{i},\mathrm{m}}(f_{\mathrm{d}})
	=
	\max_{k}\sum_{n=0}^{N-1}
	y_{\mathrm{s}}^{\mathrm{i},\mathrm{m}}[n]\{
	{y_{\mathrm{r}}^{\mathrm{i},\mathrm{m}}}[n-k]\}^{*}
	e^{-j2\pi f_{\mathrm{d}} n T_s},
\end{align}
where ${\{.\}^*}$ denotes the complex conjugate and $k$ the index of the delay bin.

Hence, a Doppler frequency of $f_{\mathrm{d}}$ Hz can be found if $R^{\mathrm{i,m}}(f_{\mathrm{d}})>\beta^{\mathrm{i,m}}(f_{\mathrm{d}})$, where $\beta^{\mathrm{i,m}}(f_{\mathrm{d}})$ is the detection threshold. According to \cite{CFAR}, 

\begin{align}
	\beta^{\mathrm{i,m}}(f_{\mathrm{d}})
	=
	\frac{1}{2W+1}
	\sum_{p=-W}^{W}
	R^{\mathrm{i},\mathrm{m}}(f_{\mathrm{d}} + p\Delta f),
\end{align}
where $W$ is the half length of training cells, and the $\Delta f$ is the resolution of Doppler frequency.

Due to the sidelobes of surveillance beams, the Doppler effect of the mobile blocker may be detected by multiple beams. Hence, the estimated direction of the mobile blocker is the one maximizing the CAF. Denote $\hat{f}_{\mathrm{i}}$ and $\hat{\mathrm{m}}_\mathrm{i}$ as the estimated Doppler frequency and direction of the mobile target in the i-th sweeping period, respectively. They are given by
\begin{align}
	\mathbf{g}_{\mathrm{i}} = [\hat{f}_{\mathrm{i}},\hat{\mathrm{m}}_\mathrm{i}]
	=
	\mathop{\arg\max}_{(f_{\mathrm{d}},\mathrm{m})}
	\left\{R^{\mathrm{i},\mathrm{m}}(f_{\mathrm{d}})\big|	
	R^{\mathrm{i},\mathrm{m}}(f_{\mathrm{d}})\geq\beta^{\mathrm{i,m}}(f_{\mathrm{d}})\right\},
\end{align}
where $\mathbf{g}_{\mathrm{i}}$ denotes the sensed feature vector in i-th sweeping period.

\section{Blockage Prediction}

In this section, the trajectory of the mobile blocker will be estimated via the detected Doppler frequencies and directions in the previous K sweeping periods, denoted as $\mathbf{G}_\mathrm{K}=[\mathbf{g}_1^{\mathsf{T}},\mathbf{g}_2^{\mathsf{T}},\ldots,\mathbf{g}_{\mathrm{K}}^{\mathsf{T}}]$. Specifically, it is assumed that the mobile blocker is moving with constant velocity in the region close to the LoS path and the distance between the transmitter and the receiver, denoted as $d$, has been estimated via the existing methods, e.g., multi-tone ranging \cite{Multi_Tone}. We first derive the AoAs and Doppler frequencies of the scattered signals from the mobile blocker versus the period index for arbitrary initial positions, motion velocities and directions, which is referred to as the motion sensing model $\Omega_K$. Then the most likely initial position, motion velocity and direction can be estimated by matching the estimated feature matrix $\mathbf{G}_\mathrm{K}$ with the motion sensing model $\Omega_K$.

\subsection{Doppler and AoA Models of Blocker Motion}
\begin{figure}[htbp]
	\centering
	\includegraphics[width=0.8\columnwidth]{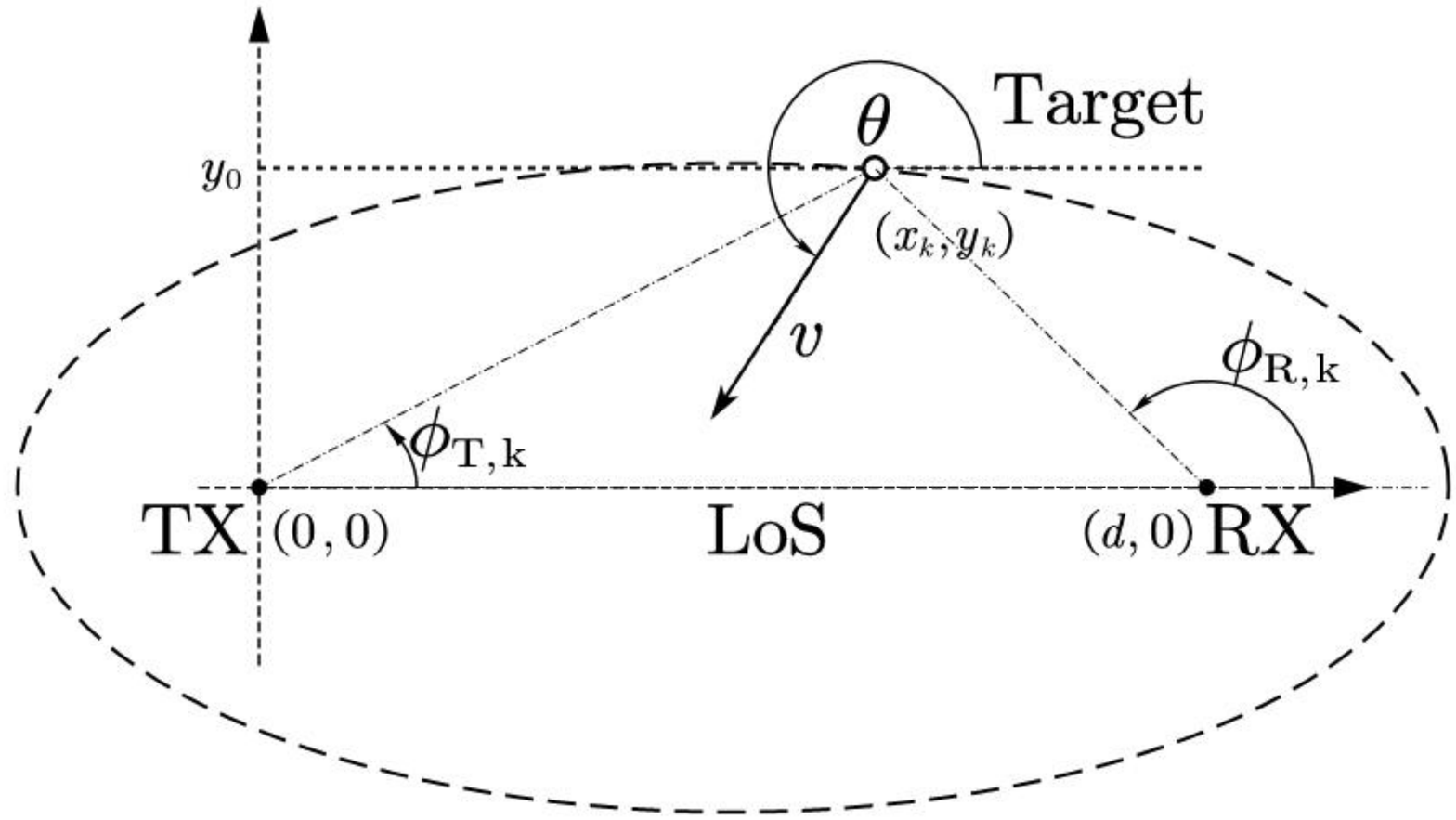}
	\caption{Illustration of the coordinate system.}
	\label{fig:Doppler}
\end{figure}

Without loss of generality, it is assumed that the transmitter and the receiver are at the origin and $(d,0)$ of the coordinate system, respectively, as illustrated in Fig. \ref{fig:Doppler}.

Suppose the blocker is moving with the velocity $v$ along the direction $\theta$, as illustrated in Fig. 2. Denote its position in the $k$-th sweeping period as $(x_{k},y_{k})$, then we have
\begin{align}
	\label{eqn:x_k}
	\begin{cases}
		x_{k}=x_{k-1}+v\mathrm{T}_{\mathrm{d}}\cos\theta
		\\
		y_{k} = y_{k-1}+v\mathrm{T}_{\mathrm{d}}\sin\theta.
	\end{cases}
\end{align}

Moreover, the AoA $\phi_{\mathrm{R,k}}$ and Doppler frequency of the blocker sensed by the receiver in the $k$-th sweeping period are given by
\begin{align}
	\label{eqn:aoa_aod}
	\phi_{\mathrm{R,k}}=\arctan(\frac{y_{k}}{x_{k}-d}),
\end{align}
and 
\begin{align}
	\begin{aligned}
	f_{d,k}
	=
	 & -2\frac{f_{\mathrm{c}}}{c}v
	\cos\left(\theta-\frac{\phi_{\mathrm{T,k}}+\phi_{\mathrm{R,k}}}{2}\right) \\
	 & \times
	\cos\left(\frac{\phi_{\mathrm{T,k}}-\phi_{\mathrm{R,k}}}{2}\right),
	\end{aligned}
	\label{eqn:fd}
\end{align}
respectively, where $\phi_{\mathrm{T,k}}=\arctan(\frac{y_{k}}{x_{k}})$, $f_{\mathrm{c}}$ and $c$ are the angle-of-departure (AoD), carrier frequency and light speed, respectively. As a result, the aggregation of Doppler frequencies and AoAs of the scattered signals from the mobile blocker in all the K sweeping periods, given the initial position $(x_{1},y_{1})$, velocity $v$ and moving direction $\theta$,  can be expressed as  
\begin{align}
	\Omega_{\mathrm{K}}(x_{1},y_{1},v,\theta) = \left[\mathbf{F}_\mathrm{d},\Phi_\mathrm{R}\right]^\mathsf{T},
\end{align}
where $\mathbf{F}_\mathrm{d} = [f_{d,1},f_{d,2},\ldots,f_{d,\mathrm{K}}]^\mathsf{T}$ and $\Phi_\mathrm{R} = [\phi_{\mathrm{R},1},\phi_{\mathrm{R},2},\ldots,\phi_{\mathrm{R},\mathrm{K}}]^\mathsf{T}$ are vectors of the Doppler frequencies and AoAs, respectively.



\subsection{Trajectory Estimation and Blockage Prediction}
The trajectory of the mobile blocker can be estimated by matching the estimated feature matrix $\mathbf{G}_\mathrm{K}$ and motion sensing model $\Omega_{\mathrm{K}}$ with appropriate initial position, motion velocity and direction, as follow. 
\begin{align}
	(\hat{x}_1,\hat{y}_1,\hat{v},\hat{\theta})
	=\mathop{\arg\min}_{(x_1,y_1,v,\theta)}
	\left\|\mathbf{w}^{\mathsf{T}}(\mathbf{G}_\mathrm{K} - \Omega_{\mathrm{K}}(x_1,y_1,v,\theta))\right\|^2,
	\label{eqn:feature}
\end{align}
where $\mathbf{w}=[w_{1},w_{2}]^{\mathsf{T}}$, $w_{1}$ and $w_{2}$ are constant weights on the Doppler frequency and AoA in trajectory estimation, respectively.
The solution can be derived by an exhaustive search of all possible combinations of initial position, motion velocity and direction.

Based on the estimated initial position, motion velocity and direction, the LoS blockage can be predicted as follows.
\begin{align}
	\mathrm{B}(\hat{x}_1,\hat{y}_1,\hat{v},\hat{\theta})
	= \left\{
	\begin{aligned}
		1 \qquad & \mathrm{if} \ \hat{\phi}_{\mathrm{T,1}} \leq 2\pi - \hat{\theta} \leq \pi - \hat{\phi}_{\mathrm{R,1}} \\
		0 \qquad & \mathrm{otherwise}
	\end{aligned}
	\right. ,
\end{align}
where $\hat{\phi}_{\mathrm{T,1}}$ and $\hat{\phi}_{\mathrm{R,1}}$ denote the estimated AoD and AoA in the first sensing period, respectively. Moreover, $\mathrm{B}(\hat{x}_1,\hat{y}_1,\hat{v},\hat{\theta})$ is an indicator whose value is 1 for LoS blockage, and 0 otherwise.

Note that the solution of trajectory estimation in \eqref{eqn:feature} may not be unique. Therefore, the LoS blockage is random and its probability shall be evaluated. 
Let $\mathcal{G}$ be the set of tuples of initial position, velocity and direction minimizing \eqref{eqn:feature}, and $\mathcal{g}_b= \{(\hat{x}_1,\hat{y}_1,\hat{v},\hat{\theta})\in \mathcal{G}| \mathrm{B}(\hat{x}_1,\hat{y}_1,\hat{v},\hat{\theta})=1\}$ be the subset of tuples leading to LoS blockage. Then the probability of LoS blockage $\mathrm{P}_{\mathrm{B}}$ can be calculated by
\begin{align}
	\mathrm{P}_{\mathrm{B}} = \frac{|\mathcal{g}_b|}{|\mathcal{G}|}.
	\label{eqn:Pb}
\end{align}

\section{Experiments and Discussion}
\begin{figure}[htbp]
	\centering
	\includegraphics[width=1\columnwidth]{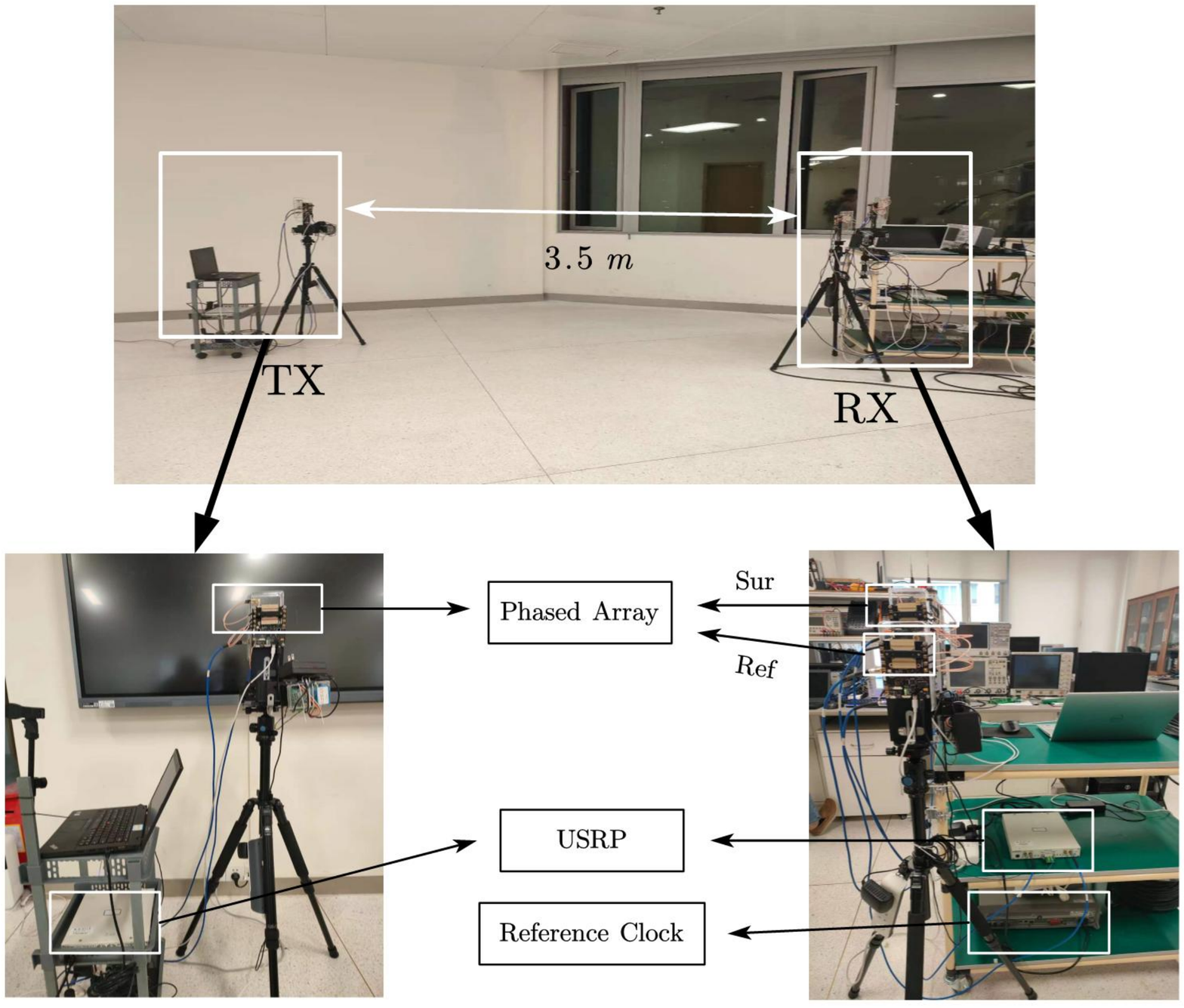}
	\caption{The mmAlert of mmWave blockage prediction system.}
	\label{fig:System}
\end{figure}

The mmAlert was implemented via software-defined radio (SDR) and mmWave phased array as shown in Fig. \ref{fig:System}. The transmitter is composed of one NI USRP-2954R and one Sivers 60 GHz phased array. The transmit signal consists of a training sequence and OFDM-modulated data payload. The signal bandwidth is 10 MHz and the beam width is 90°.
At the receiver, two Sivers 60 GHz phased arrays are connected with one SDR.
The beamwidths of the reference beam and the surveillance beam are both 10°. The surveillance beam is switched periodically among $\mathrm{M} = 4$ directions, which are 40°, 27°, 18° and 10°. The sensing time of one direction is $\mathrm{T}_{\mathrm{b}} = 25$ ms and the duration of one sweeping period is $\mathrm{T}_{\mathrm{d}}= 100$ ms.

\begin{figure}[htbp]
	\centering
	\includegraphics[width=0.8\columnwidth]{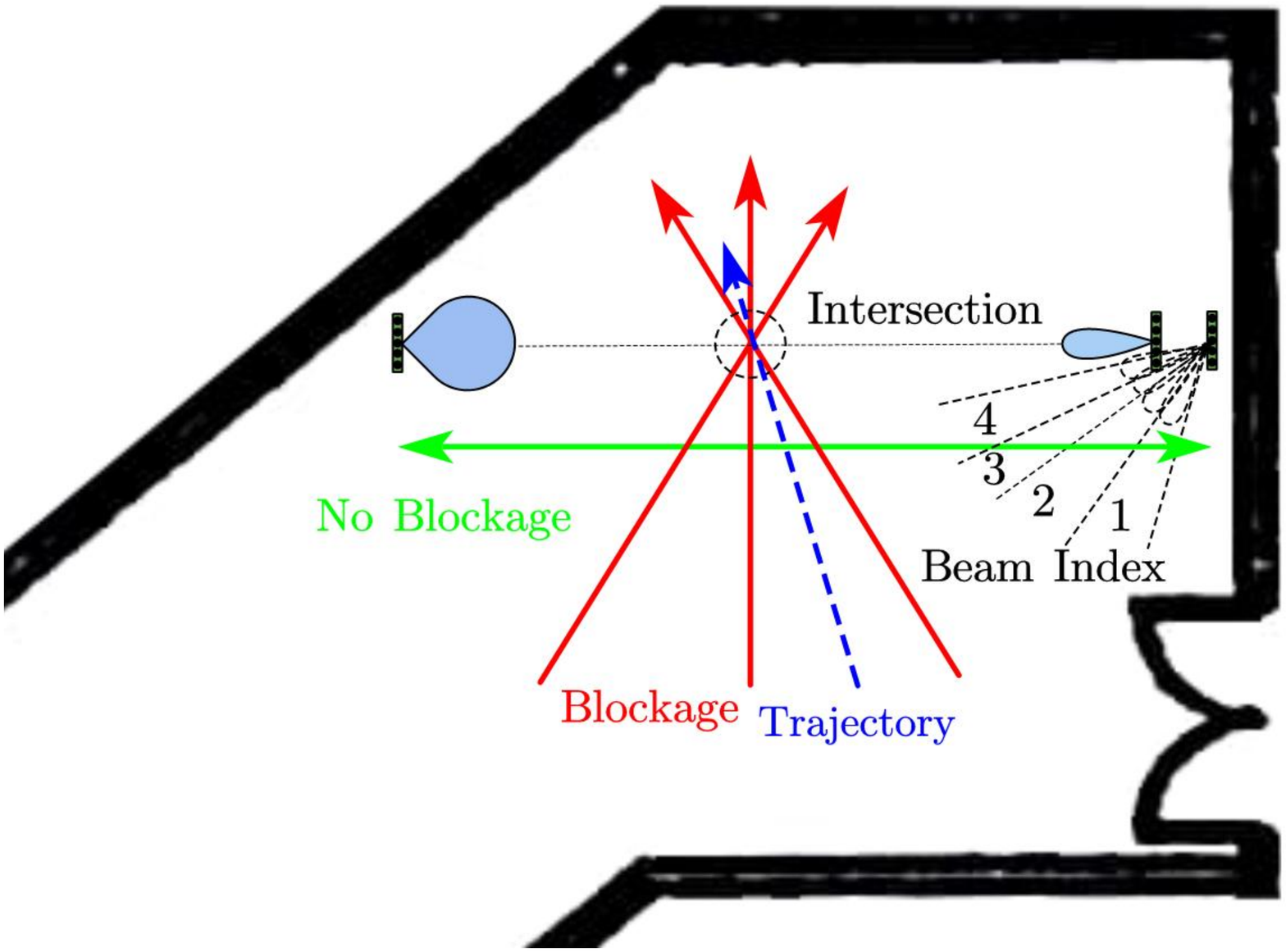}
	\caption{The experiment scenario in an open corridor.}
	\label{fig:Scenario}
\end{figure}

As shown in Fig. \ref{fig:Scenario}, the experiment is conducted in an indoor environment. 
All of the phased arrays are placed at a height of 1.35 m. The distance between the transmitter and the receiver is 3.5 m. 
In the experiment, one person walks with a constant velocity close to the LoS path. As illustrated in Fig. \ref{fig:Scenario}, two types of trajectories are tested: the green trajectories will not block the LoS path, but the red ones will do. 

\subsection{Doppler and AoA Estimation}

\begin{figure}[htbp]
	\centering
	\subfloat[]{
		\includegraphics[width=0.48\columnwidth]{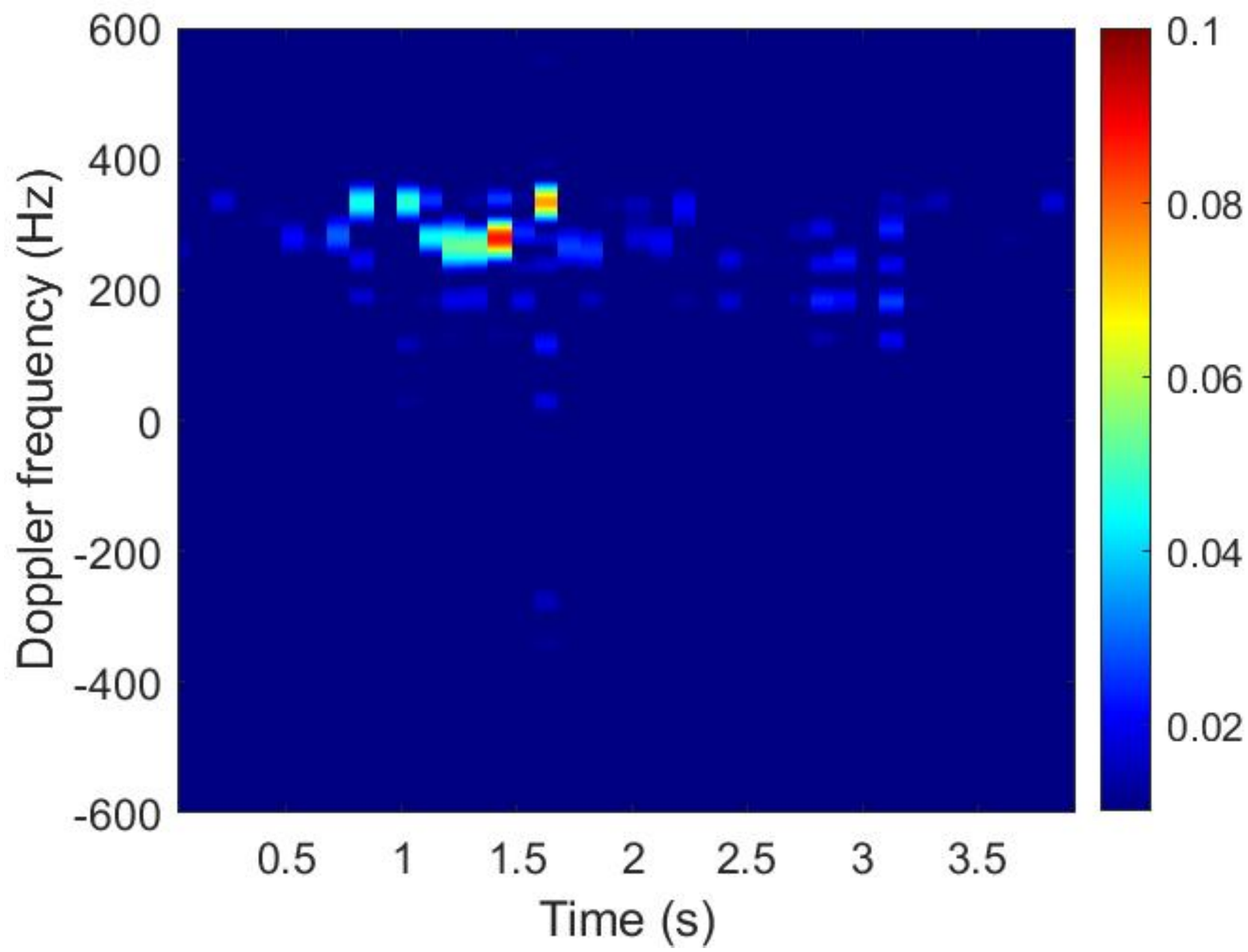}
		\label{a}
	}
	\subfloat[]{
		\includegraphics[width=0.48\columnwidth]{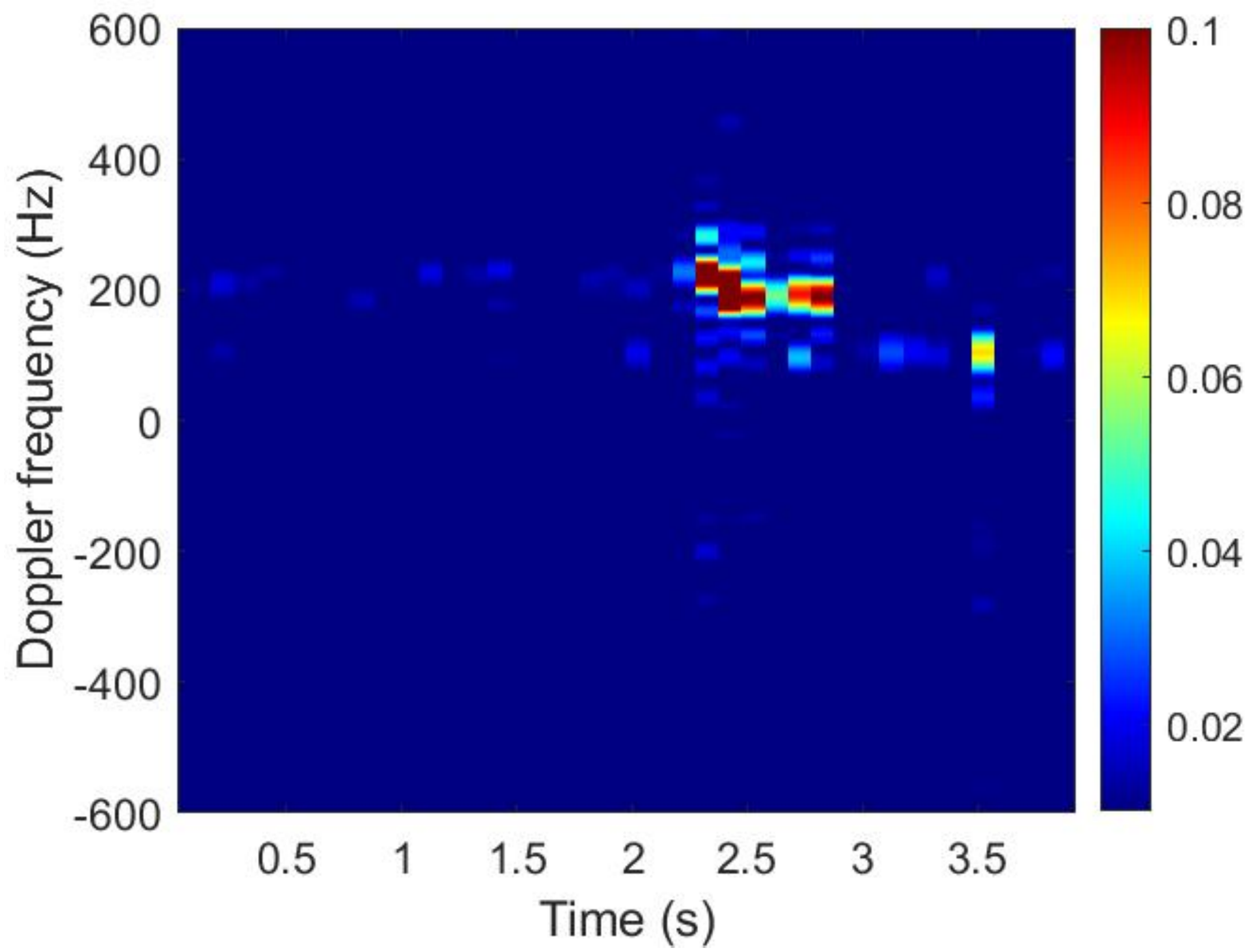}
		\label{b}
	}
	\\
	\subfloat[]{
		\includegraphics[width=0.48\columnwidth]{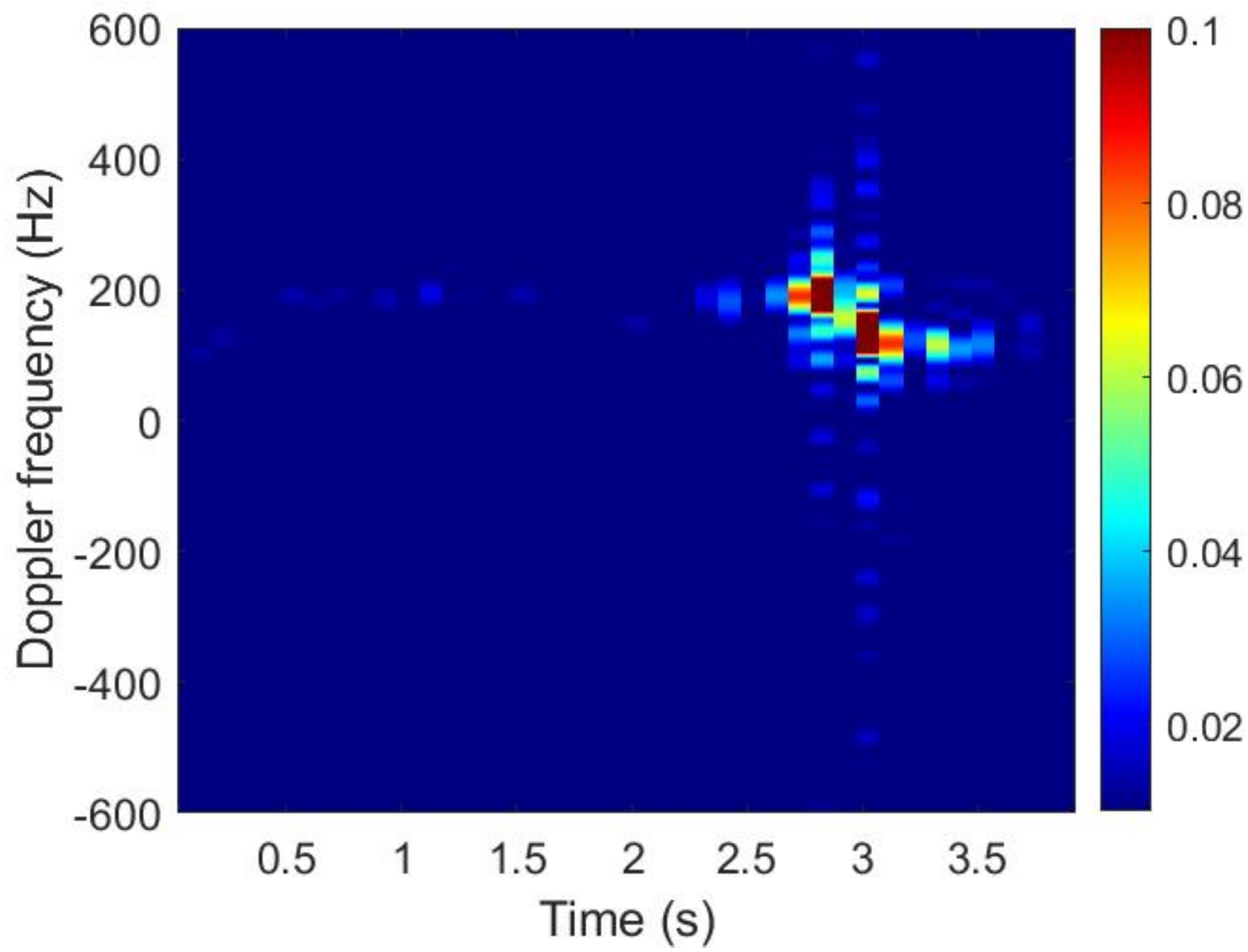}
	}
	\subfloat[]{
		\includegraphics[width=0.48\columnwidth]{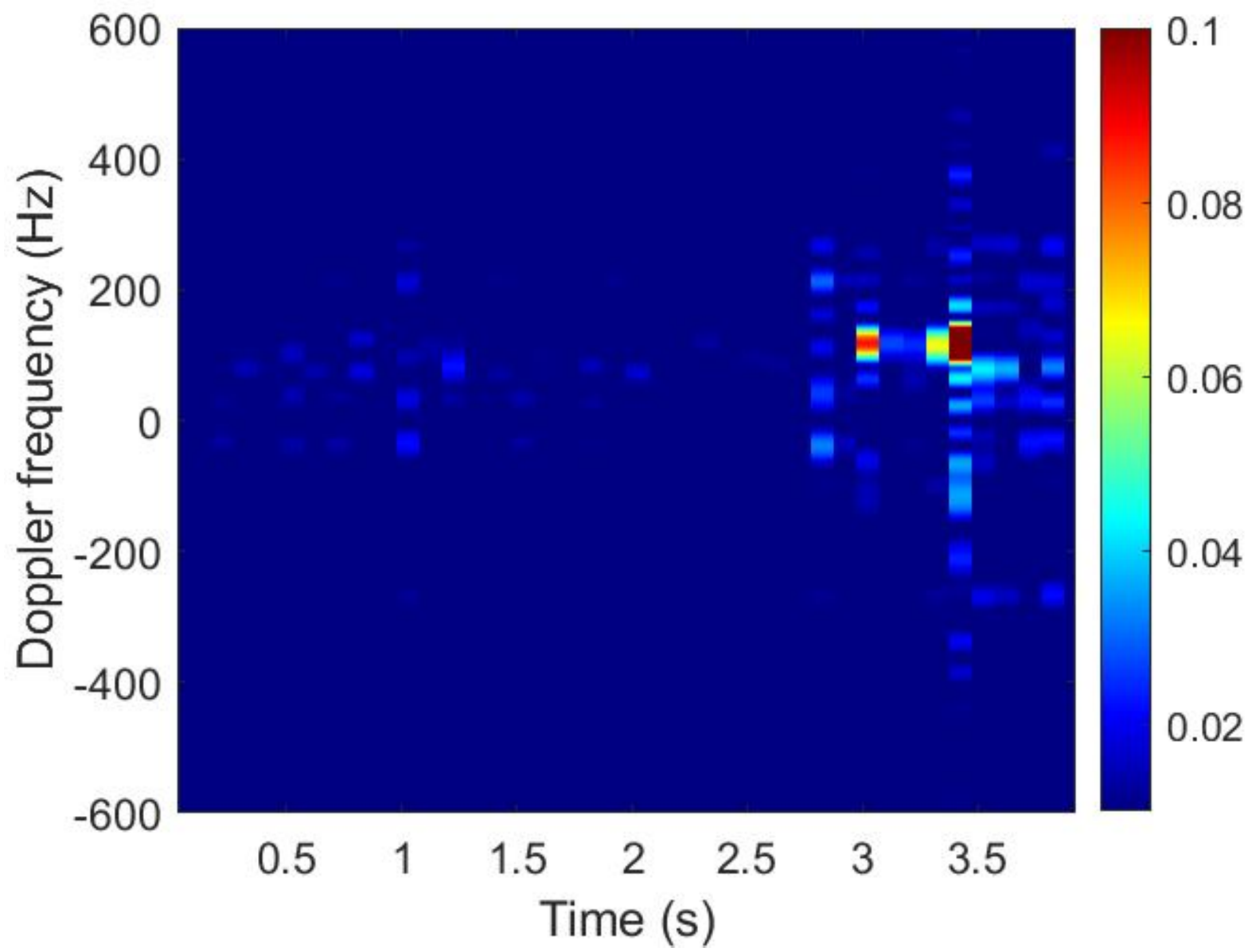}
		\label{d}
	}

	\caption{The (a) to (d) show the time-Doppler spectrograms of the beams 1 to 4, respectively.}
	\label{fig:Time-Doppler}
\end{figure}

The time-Doppler spectrograms of all 4 surveillance beams, illustrating the Doppler frequency versus time, in a trail is shown in Fig. \ref{fig:Time-Doppler}. 
The total sensing duration is 4 s, which consists of about 40 sweeping periods. In this trial, the person walks with a velocity of 1 m/s. The trajectory crosses the LoS path at the time instance 3.8 s, as shown by the blue line of in Fig. \ref{fig:Scenario}.
It can be observed that the Doppler effect of the walking person is captured by the four surveillance beams in the time intervals 0.8 to 1.7 s, 2.2 to 2.7 s, 2.7 to 3.3 s and 3.3 to 3.5 s, respectively. The Doppler frequency decreases when the person approaches the LoS path. Moreover, due to the existence of sidelobes, the Doppler effect can be found even when the person is not within the coverage of the mainlobe.  

\begin{figure}[htbp]
	\centering
	\includegraphics[width=0.8\columnwidth]{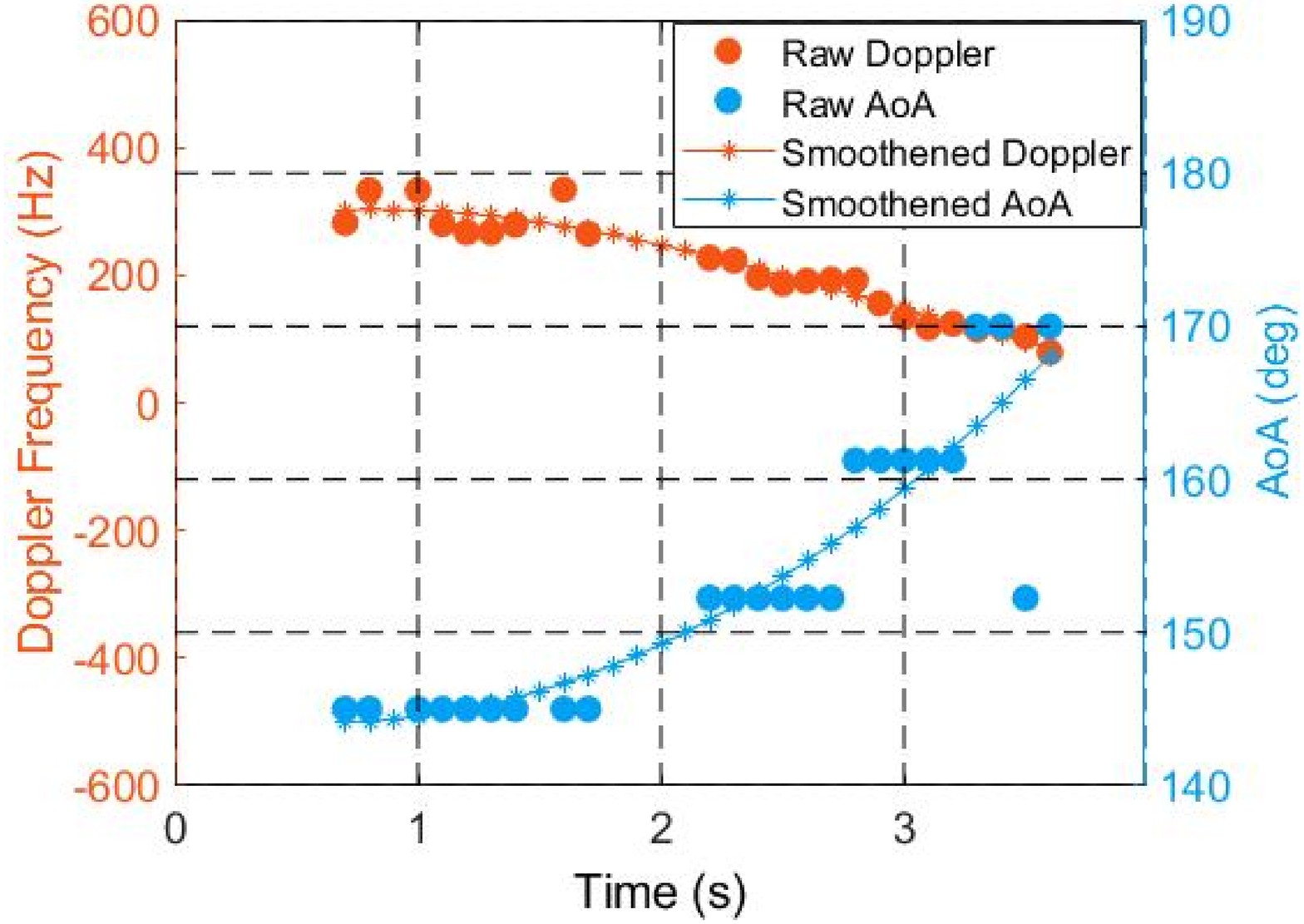}
	\caption{The estimated Doppler frequencies and AoAs in total sensing duration.}
	\label{fig:Fitted}
\end{figure}

Integrating the spectrograms of all four surveillance beams, the estimated Doppler frequency and AoA versus time of a trail is shown in Fig. \ref{fig:Fitted}, where the estimated Doppler frequency and AoA are smoothened via polynominals (the red and blue dots are the raw estimation of Doppler frequency and AoA, respectively). Note that the resolution of AoA can be improved if more surveillance beams with smaller beam width are adopted. Moreover, miss detection and false alarm can be found in the estimation. For example, no target is found at the time interval from 1.7 to 2.2 s, and a wrong AoA is detected at 3.5 s. In order to eliminate the effect of false alarm and miss detection, the curve-fitting approach is adopted to match the measured Doppler frequencies and AoAs with smooth polynomial curves. Hence, the fitted traces of Doppler frequencies and AoAs, instead of the raw estimation, are used in the detection of initial position, motion velocity and direction in \eqref{eqn:feature}. 

\subsection{Accuracy of Blockage Prediction}
In this part, the blockage prediction accuracy is investigated with 100 samples with different directions and velocities, where 80 trajectories would lead to blockage. The blockage prediction of these 100 samples is conducted by the method in Section \ref{section:signal processing}, where different sensing durations are tested.

\begin{figure}[htbp]
	\centering
	\subfloat[]{\includegraphics[width=0.48\columnwidth]{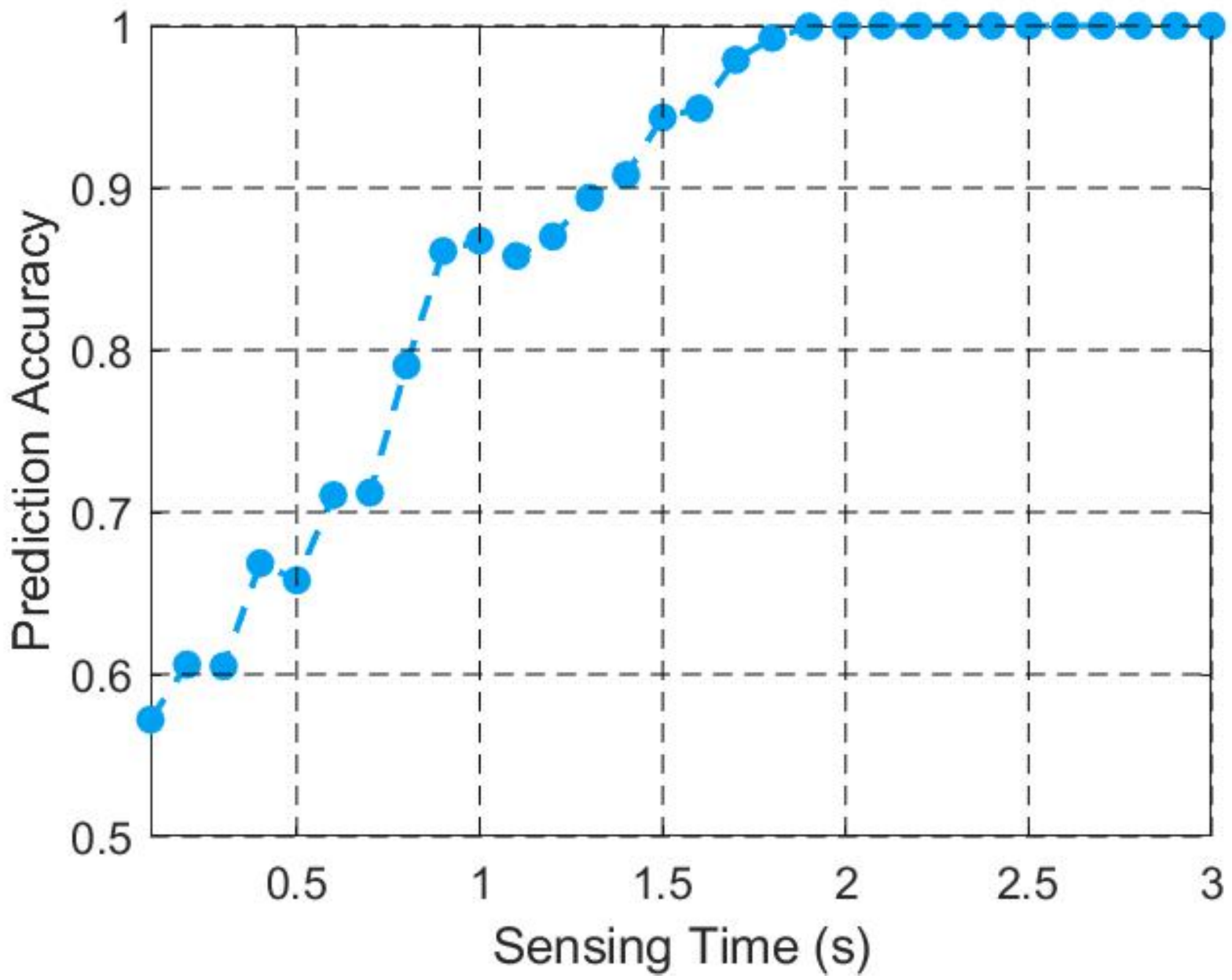}
	\label{fig:Observation}}
	\subfloat[]{\includegraphics[width=0.48\columnwidth]{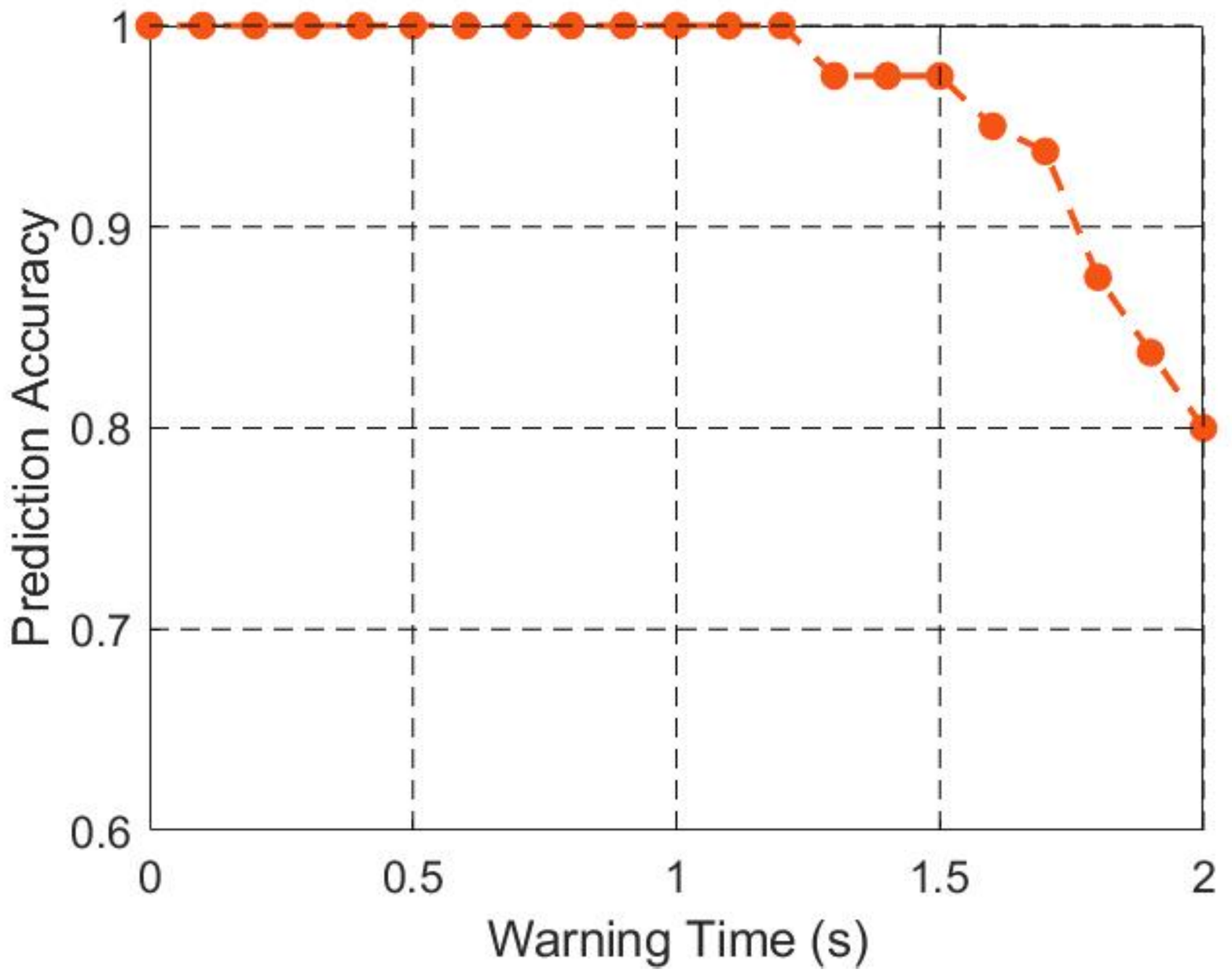}
	\label{fig:Blockage Prediction}}
	\caption{The (a) is the prediction probability versus the sensing duration and the (b) is the blockage detection accuracy versus the warning time.}
\end{figure}

In Fig. \ref{fig:Observation}, the prediction accuracy versus the sensing duration is plotted. 
In this figure, a trajectory estimation with blockage probability in \eqref{eqn:Pb} above 90\% is believed to block the LoS path soon.
The prediction accuracy refers to the percentage of tested trajectories whose blockage prediction is correct. The larger the sensing duration, the more sweeping periods are used to estimate the trajectories. It can be observed that more than 90\% of the estimated trajectory is correctly predicted when the sensing duration exceeds 1.4 s.

In order to show the blockage warning time provided by the mmAlert system, the blockage prediction accuracy versus the warning time is provided in Fig. \ref{fig:Blockage Prediction}. It can be observed that 100\% blockage in future 1.2 s can be detected with an accuracy of 100\%, and the prediction accuracy is still above 90\% for 1.8 s warning time. This demonstrates that the mmAlert system could provide sufficient warning time for backup (NLoS) path detection and link setup.

\section{Conclusion}
In this letter, the passive sensing technique is integrated with a mmWave communication system for LoS blockage detection. Specifically, the surveillance beam at the receiver periodically sweeps the region close to the LoS path, such that the Doppler frequency and AoA of the scattered signal from the mobile blocker can be detected. By tracking the variation of the Doppler frequencies and AoAs, the trajectory of the mobile blocker can be estimated, and hence, the proposed mmAlert system is able to predict whether the mobile blocker would block the LoS path. It is demonstrated by experiments that the prediction accuracy of mmAlert system is high and the sufficient warning time can be provided.

\bibliographystyle{IEEEtran}
\bibliography{Mmwave_Blockage}
\end{document}